\documentclass[10pt]{article}
\usepackage{graphics,epsfig}
\usepackage[]{graphicx}
\usepackage{latexsym}

\font\smallroman=cmr9

\begin{document}
\title{\bf The Contextual Character of \\ Modal Interpretations of Quantum
Mechanics}
\author{{\sc Graciela
Domenech}\thanks{%
Fellow of the Consejo Nacional de Investigaciones Cient\'{\i}ficas
y T\'ecnicas (CONICET)} $^{1, 3}$,\  \  {\sc Hector Freytes}
$^{2}$\ {\sc and} \ {\sc Christian de Ronde} $^{3, 4}$}
\date{}
\maketitle

\begin{center}

\begin{small}
1. Instituto de Astronom\'{\i}a y F\'{\i}sica del Espacio (IAFE)\\
Casilla de Correo 67, Sucursal 28, 1428 Buenos Aires, Argentina\\
2. Dipartimento di Scienze e Pedagogiche e Filosofiche -
Universita degli Studi di Cagliari \\ Via Is Mirrionis 1, 09123,
Cagliari - Italia\\ 3. Center Leo Apostel (CLEA)\\ 4. Foundations
of the Exact Sciences (FUND) \\ Brussels Free University -
Krijgskundestraat 33, 1160 Brussels - Belgium
\end{small}
\end{center}

\begin{abstract}
\noindent In this article we discuss the contextual character of
quantum mechanics in the framework of modal interpretations. We
investigate its historical origin and relate contemporary modal
interpretations to those proposed by M. Born and W. Heisenberg. We
present then a general characterization of what we consider to be
a modal interpretation. Following previous papers in which we have
introduced modalities in the Kochen-Specker theorem, we
investigate the consequences of these theorems in relation to the
modal interpretations of quantum mechanics.
\end{abstract}
\bigskip
\noindent

\newtheorem{theo}{Theorem}[section]

\newtheorem{definition}[theo]{Definition}

\newtheorem{lem}[theo]{Lemma}

\newtheorem{prop}[theo]{Proposition}

\newtheorem{coro}[theo]{Corollary}

\newtheorem{exam}[theo]{Example}

\newtheorem{rema}[theo]{Remark}{\hspace*{4mm}}

\newtheorem{example}[theo]{Example}

\newcommand{\proof}{\noindent {\em Proof:\/}{\hspace*{4mm}}}

\newcommand{\qed}{\hfill$\Box$}

\newcommand{\ninv}{\mathord{\sim}} 

\section*{Introduction}

Modal propositions refer to possibility and necessity, about what
`must be' and what `may be' the case.  If one thinks about how
things are in the actual world, then one may also think of how
things might have been in an alternative, non actual but possible,
state of affairs. Modalities were introduced in physics by
statistical mechanics in order to take into account possible
configurations of physical systems regardless of what is actually
the case. In contemporary physics, the development of quantum
theories has thrown new light to the problem of the relation
between possible and actual. We intend to analyze the r\^{o}le of
modalities in quantum mechanics, precisely in what sense we may
say that a system possibly possesses a property. We also study
whether the addition of modal propositions to the lattice of
actual propositions would circumvent the contextual character of
the quantum discourse about properties.

The paper is organized in five sections. In section 1, we study
the modal character of early interpretations of quantum mechanics.
In section 2, we discuss the contextual character of the theory.
In section 3, we outline a brief review of contemporary modal
interpretations. We then present, in section 4, a general
characterization of these interpretations. In section 5 we analyze
their contextual character. Finally, we present a brief
discussion.

\section{Early Modal Interpretations of Quantum Mechanics}

The first consistent formalization of the methods applied to the
study of quantum phenomena was developed in the form of `matrix
mechanics' first under the ideas of W. Heisenberg, and then in
collaboration with M. Born and P. Jordan, during the second half
of 1925 (Heisenberg, 1925; Born and Jordan, 1925; Born, Heisenberg
and Jordan, 1926):

{\smallroman
\begin{quotation}
``[...] in the summer of 1925 it led to a mathematical formalism
called matrix mechanics or, more generally, quantum mechanics. The
equations of motion of Newtonian mechanics were replaced by
similar equations between matrices; it was a strange experience to
find that many of the old results of Newtonian mechanics, like
conservation of energy, etc, could be derived also in the new
scheme. Later the investigations of Born, Jordan and Dirac showed
that the matrices representing position and momentum of the
electron did not commute. This latter fact demonstrated clearly
the essential difference between quantum mechanics and classical
mechanics.'' W. Heisenber (1958, p. 41)
\end{quotation}}

In January 1926 E. Schr\"odinger completed the first part of his
apparently quite different approach to the same problem in terms
of eigenfunctions of a differential equation, which is known as
`wave mechanics' (Schr\"odinger, 1926a). Schr\"odinger had read
Einstein's paper on the quantum theory of the ideal gas. He
recognized later it was these ideas, due to Einstein and de
Broglie, which allowed him to develop his own theory (Jammer,
1966, p. 257). By March 1926, Schr\"odinger interpreted the wave
function $\psi$ as a `mechanical field scalar' ({\it mechanischer
Feldskalar}) in connection with the classical theory of
electromagnetic radiation. Following this idea, the eigenvalues
where taken as frequency values (Schr\"odinger, 1926c). The
quantum postulate was then conceived in connection with resonance
phenomena, making possible to reject the existence of discrete
energy levels and quantum jumps. Schr\"odinger's wave theory was
well received by the orthodox physics community which wanted to
put an end to the deviations of the atomic theory from the
classical conception of natural sciences developed from Newton to
Einstein. At the time Heisenberg was quite unhappy about this
turn:

{\smallroman
\begin{quotation}
``In July [1926] I visited my parents in M\"unich and on this
occasion I heard a lecture given by Schr\"{o}dinger for the
physicists in M\"unich about his work on wave mechanics. It was
thus that I first became acquainted with the interpretation
Schr\"{o}dinger wanted to give his mathematical formalism of wave
mechanics, and I was disturbed about the confusion with which I
believed this would burden atomic theory. Unfortunately, nothing
came from my attempt during the discussion to put things in order.
My argument that one could not even understand Plank's radiation
law on the basis of Schr\"{o}dinger's interpretation convinced no
one. Wilhem Wien, who held the chair of experimental physics at
the University of M\"unich, answered rather sharply that one must
put an end to quantum jumps and the whole atomic mysticism, and
the difficulties I had mentioned would certainly soon be solved by
Schr\"{o}dinger.'' W. Heisenberg (quoted from Wheeler and Zurek,
1983, p. 56)
\end{quotation}}

Schr\"odinger related the matter field and the electromagnetic
emission or absorption of atomic systems in the famous paper in
which he showed the equivalence between matrix and wave mechanics
(Schr\"odinger, 1926b). To do so, he first linked the wave
function $\psi$ to a space density of the electrical charge given
by the real part of $\psi\partial\psi^{*}/\partial t$, where
$\psi^{*}$ denotes the complex conjugate of $\psi$, correcting it
later to $\psi\psi^{*}$. This interpretation, in terms of a field,
had many difficult features and internal inconsistencies, which
would only be solved, later on, by M. Born.

\subsection{Born's Modal Interpretation of the Quantum Wave
Function}

At the same time of Schr\"odinger's demonstration of the
equivalence of matrix and wave mechanics, Born presented his
probabilistic interpretation of the wave function. As commented by
M. Jammer:

{\smallroman
\begin{quotation}
``Almost simultaneously with the appearance of Schr\"odinger's
fourth communication, a new interpretation of the $\psi$-function
was published that had far-reaching consequences for modern
physics not only from the purely technical point of view but also
with respect to the philosophical significance of its content.
Only four days after Schr\"odinger's concluding contribution had
been sent to the editor of the {\it Annalen der Physik} the
publishers of the {\it Zeitschrift f\"ur Phisik} received a paper,
less than five pages long, titled `On the Quantum Mechanics of
Collision Processes' in which Max Born proposed, for the first
time, a probabilistic interpretation of the wave function implying
thereby that microphysics must be considered a probabilistic
theory.'' M. Jammer (1974, p. 38)
\end{quotation}}

In his original article, Born begins by explicitly characterizing
quantum mechanics as a modal theory, emphasizing its probabilistic
character but stressing at the same time the indeterministic
element present in the theory:

{\smallroman
\begin{quotation}
``Schr\"{o}dinger's quantum mechanics [therefore] gives quite a
definite answer to the question of the effect of the collision;
but there is no question of any causal description. One gets no
answer to the question, `what is the state after the collision'
but only to the question, `how probable is a specified outcome of
the collision'.

Here the whole problem of determinism comes up. {\it From the
standpoint of our quantum mechanics there is no quantity which in
any individual case causally fixes the consequence of the
collision; but also experimentally we have so far no reason to
believe that there are some inner properties of the atom which
condition a definite outcome for the collision.} [...] I myself am
inclined to give up determinism in the world of the atoms. But
that is a philosophical question for which physical arguments
alone are not decisive.'' M. Born (1926, quoted from Wheeler and
Zurek, 1983, p. 57, emphasis added)
\end{quotation}}

\noindent It soon became evident that the interpretation of
modality in the new theory departed from its use in classical
statistical mechanics as lack of knowledge. In the years to come
this would remain one of the main themes regarding the physical
interpretation of quantum mechanics.

{\smallroman
\begin{quotation}
``It was wave or quantum mechanics that was first able to assert
{\it the existence of primary probabilities in the laws of
nature}, which accordingly do not admit of reduction to
deterministic natural laws by auxiliary hypotheses, as do for
example the thermodynamic probabilities of classical physics. This
revolutionary consequence is regarded as irrevocable by the great
majority of modern theoretical physicists –-primarily by M. Born,
W. Heisenberg and N. Bohr, with whom I also associate myself."  W.
Pauli (1994, p. 46)\end{quotation}}

\subsection{Heisenberg's Modal Interpretation}

Even though Born's interpretation of the wave function might seem
to solve the interpretational problems by making reference to
probabilities, it became immediately clear that this resource
would not avoid the main difficulties. As Schr\"{o}dinger noted in
a letter to Einstein dated November 18, 1950, the so called
`quantum probabilities' departed from what was commonly understood
as {\it probability}:

{\smallroman
\begin{quotation}
``It seems to me that the concept of probability is terribly
mishandled these days. Probability surely has as its substance a
statement as to whether something {\it is} or {\it is not} the
case --an uncertain statement, to be sure. But nevertheless it has
meaning only if one is indeed convinced that the something in
question quite definitely {\it is} or {\it is not} the case. A
probabilistic assertion presupposes the full reality of its
subject.'' E. Schr\"{o}dinger (1950, quoted from Bub, 1997, p.
115)
\end{quotation}}

\noindent Heisenberg himself was eager to repeatedly remark the
essential distance between classical and quantum probabilities:

{\smallroman
\begin{quotation}
``For each complementary statement the question [whether the atom
is in the left or the right half of the box] is not decided. But
the term `not decided' is by no means equivalent to the term `not
known'. `Not known' would mean that the atom is `really' left or
right, only we do not know where it is. But `not decided'
indicates a different situation, expressible only by a
complementary statement.'' W. Heisenberg (1958, p.
158).\end{quotation}}

\noindent His own interpretation, although not completely clear,
related possibility to the Aristotelian concept of {\it potentia}:

{\smallroman
\begin{quotation}
``[...] the paper of Bohr, Kramers and Slater revealed one
essential feature of the correct interpretation of quantum theory.
This concept of the probability wave was something entirely new in
theoretical physics since Newton. Probability in mathematics or in
statistical mechanics means a statement about our degree of
knowledge of the actual situation. In throwing dice we do not know
the fine details of the motion of our hands which determine the
fall of the dice and therefore we say that the probability for
throwing a special number is just one in six. The probability wave
function of Bohr, Kramers and Slater, however, meant more than
that; it meant a tendency for something. It was a quantitative
version of the old concept of `potentia' in Aristotelian
philosophy. It introduced something standing in the middle between
the idea of an event ant the actual event, a strange kind of
physical reality just in the middle between possibility and
reality.'' W. Heisenberg (1958, p. 42)
\end{quotation}}

It is important to remark that in Heisenberg's interpretation the
projection postulate is not considered to be a physical process as
is the case, for example, of GRW-type theories which intend to
provide a dynamical account of the collapse of the wave function
(Ghirardi, Rimini, Weber, 1985):

{\smallroman
\begin{quotation}
``The observation itself changes the probability function
discontinuously; it selects of all possible events the actual one
that has taken place. [...] When the old adage 'Natura non facit
saltus' is used to criticism of quantum theory, we can reply that
certainly our knowledge can change suddenly and that this fact
justifies the use of the term 'quantum jump'.'' W. Heisenberg
(1958, p. 54)
\end{quotation}}

Modal interpretations describe the evolution of possibilities.
However, this representation must not contradict a particular
actualization, one must be able to give an account of the path
from the possible to the actual.

{\smallroman
\begin{quotation}
``The mathematical representation of the physical process changes
discontinuously with each new measurement; the observation singles
out of a large number of possibilities one of which is the one
which has happened. The wave packet which has spread out is
replaced by a smaller one which represents the result of this
observation. As our knowledge of the system does change
discontinuously at each observation its mathematical
representation must also change discontinuously; this is to be
found in classical statistical theories as well as in the present
theory." W. Heisenberg (1949, p. 36)\end{quotation}}

From the beginning it was acknowledged that quantum theory started
with a paradox: on the one hand, it describes experiments in terms
of classical physics and, on the other, it was born from the very
critical distance taken from exactly these same concepts (the
description via the trajectory). The necessity of a new image, an
{\it anschaulich} content, was early recognized by Heisenberg, but
in order to provide such visualization one needed to understand
the structure of the theory. Classically, systems are described in
terms of their actual properties, thus, in order to create a new
image consistent with the quantum formalism, it appeared important
to study the new structure of propositions about the properties of
the system. This task was accomplished by the developments of G.
Birkhoff and J. von Neumann, in what they called quantum logic.
However, this analysis did not include modal features. The
analysis of the modal properties was always placed within the
limits of Bohr's concept of complementarity; thus, the classical
space-time description (given by the definite result, i.e. a spot
in a photographic plate) was complementary to the quantum causal
description (provided by the Schr\"odinger evolution of the
quantum wave function) in a quite definite sense:

{\smallroman
\begin{quotation}
``[...] as a geometric or kinematic description of a process
implies observation, it follows that such a description of atomic
processes necessarily precludes the exact validity of the law of
causality--and conversely. Bohr has pointed out that it is
therefore impossible to demand that both requirements be fulfilled
by the quantum theory. They represent complementary and mutually
exclusive aspects of atomic phenomena." W. Heisenberg (1949, pp.
63-64)\end{quotation}}

\noindent The relation between these descriptions, namely, the
study of the relation between the possible and the actual is the
main subject of this work. In the following section we will come
to the concept of contextuality, which rigorously summarizes the
ideas of Bohr and Heisenberg as discussed above.

\section{The Contextual Character of Quantum Mechanics}

As we have already mentioned, when discussing the interpretation
of matrix mechanics, Heisenberg could not avoid recognizing that
his formalism did not provide a consistent image (an {\it
anschaulich} content), contrary to wave mechanics which could be
interpreted in terms of a field. As noted by J. Hilgevoord and J.
Uffink (2006), the purpose of his 1927 paper was to provide
exactly this lacking feature. In this paper, he developed the
undeterminancy relations, which in turn present one of the most
striking features of quantum systems, namely the fact that exact
values cannot  be simultaneously assigned to all their quantities.
This characteristic of the new mechanics, as well as its
consequences, was clearly acknowledged by Schr\"odinger:

{\smallroman
\begin{quotation}
``[...] if I wish to ascribe to the model at each moment a
definite (merely not exactly known to me) state, or (which is the
same) to all determining parts definite (merely not exactly known
to me) numerical values, then there is no supposition as to these
numerical values to be imagined that would not conflict with some
portion of quantum theoretical assertions.'' E. Schr\"odinger
(1935, quoted from Wheeler and Zurek, 1983, p. 152, emphasis
added)
\end{quotation}}

\noindent Also P.A.M. Dirac stated in his famous book:

{\smallroman
\begin{quotation}
``The expression that an observable `has a particular value' for a
particular state is permissible in quantum mechanics in the
special case when a measurement of the observable is certain to
lead to the particular value, so that the state is an eigenstate
of the observable. It may easily be verified from the algebra
that, with this restricted meaning for an observable `having a
value', if two observables have values for a particular state,
then for this state the sum of the two observables (if the sum is
an observable) has a value equal to the sum of the values of the
two observables separately and the product of the two observables
(if this product is an observable) has a value equal to the
product of the values of the two observables separately.'' P.
Dirac (1958, p. 46).
\end{quotation}}

\noindent This last point is the requirement of the {\it
functional compatibility condition} (FUNC), to which we will
return later.

In the usual terms of quantum logic (see for example Jauch, 1968;
Piron, 1976), a property of a system is related to a subspace of
the Hilbert space of its (pure) states or, analogously, to the
projector operator onto that subspace. A physical magnitude
${\mathcal A}$ is represented by an operator $\bf A$ acting over
the state space. For bounded self-adjoint operators, conditions
for the existence of the spectral decomposition ${\bf A}=\sum_{i}
a_i {\bf P}_i=\sum_{i} a_i |a_i><a_i|$ are satisfied. The real
numbers $a_i$ are related to the outcomes of measurements of the
magnitude ${\mathcal A}$ and projectors $|a_i><a_i|$ to the
properties of the physical system. Precisely, let ${\mathcal H}$
be the Hilbert space associated to the physical system and
${\mathcal L}({\mathcal H})$ be the set of closed subspaces on
${\mathcal H}$. If we consider the set of these subspaces ordered
by inclusion, with the complement defined by orthocomplementation
and meet and join operations defined by intersection and direct
sum of subspaces, then ${\mathcal L}({\mathcal H})$ is a complete
orthomodular lattice (Maeda and Maeda, 1970). It is well known
that each self-adjoint operator $\bf A$ has associated a Boolean
sublattice $W_A$ of ${\mathcal L}({\mathcal H})$. More precisely,
$W_A$ is the Boolean algebra of projectors ${\bf P}_i$ of the
spectral decomposition. We will refer to $W_A$ as the spectral
algebra of the operator $\bf A$. Any proposition about the system
is represented by an element of ${\mathcal L}({\mathcal H})$,
which is the algebra of quantum logic introduced by Birkhoff and
von Neumann (1936).

A complete set of properties of a system that may be
simultaneously predicated allows to construct a Boolean
propositional system. This is usually referred to as a {\it
context}. In terms of operators, it is in correspondence with a
complete set of commuting observables, CSCO for short. Assigning
values (a set of their corresponding eigenvalues) to these
magnitudes poses no difficulties. But if we try to interpret
eigenvalues as the actual values of the physical properties of a
system, we are faced to all kind of no-go theorems that preclude
this possibility. Most remarkably is the Kochen-Specker (KS)
theorem that rules out the non-contextual assignment of values to
physical magnitudes (Kochen and Specker, 1967). An explicit
statement of the KS theorem reads (Held, 2003):

\begin{theo}\label{c2}
Let ${\cal H}$ be a Hilbert space of dimension greater than 2 of
the states of the system and $M$ be a set of observables,
represented  by operators on ${\cal H}$. Then, the following two
assumptions are contradictory:
\begin{enumerate}
\item
All members of $M$ simultaneously have values, i.e. are
unambiguously mapped onto real unique numbers (designated, for
observables $\bf A$, $\bf B$, $\bf C$, ... by $v({\bf A})$,
$v({\bf B})$, $v({\bf C})$, ...).
\item
Values of observables conform to the following constrains:
\subitem If $\bf A$, $\bf B$, $\bf C$ are all compatible and ${\bf
C} = {\bf A}+{\bf B}$, then $v({\bf C}) = v({\bf A})+v({\bf B})$;
\subitem if $\bf A$, $\bf B$, $\bf C$ are all compatible and ${\bf
C} = {\bf A}·{\bf B}$, then $v({\bf C}) = v({\bf A})·v({\bf
B})$.\qed
\end{enumerate}
\end{theo}

As we have stated in (Domenech and Freytes, 2005), KS theorem may
be expressed in terms of families of Boolean homomorphisms.
Assigning values to a physical quantity ${\cal A}$ is equivalent
to establishing a Boolean homomorphism $v: W_A \rightarrow {\bf
2}$ (Isham and Butterfield, 1998), being ${\bf 2}$ the two
elements Boolean algebra. Thus, we may define a {\it global
valuation} over ${\mathcal L}({\mathcal H})$ as the family of
Boolean homomorphisms $(v_i: W_i \rightarrow {\bf 2})_{i\in I}$
such that $v_i\mid W_i \cap W_j = v_j\mid W_i \cap W_j$ for each
$i,j \in I$, being $(W_i)_{i\in I}$ the family of Boolean
sublattices of ${\mathcal L}({\mathcal H})$. This global valuation
would give the values of all magnitudes at the same time
maintaining a {\it compatibility condition} in the sense that
whenever two magnitudes shear one or more projectors, the values
assigned to those projectors are the same from every context. But
KS theorem assures that we cannot assign real numbers pertaining
to their spectra to the operators in such a way to satisfy $item\
2$ of \ref{c2}, i.e., FUNC, the expression of the `natural'
requirement asked by Dirac, as we have mentioned above. KS theorem
expresses the fact that, if we demand a valuation to satisfy FUNC,
then it is forbidden to define it in a non-contextual fashion for
subsets of quantities represented by commuting operators. In the
algebraic terms of the previous definition, KS theorem reads:

\begin{theo}\label{CS3}
If $\mathcal{H}$ is a Hilbert space such that $dim({\cal H}) > 2$,
then a global valuation over ${\mathcal L}({\mathcal H})$ is not
possible.\qed
\end{theo}

Of course contextual valuations allow us to refer to different
sets of actual properties of the system which define its state in
each case. Algebraically, a {\it contextual valuation} is a
Boolean valuation over one chosen spectral algebra. In classical
particle mechanics it is possible to define a Boolean valuation of
all propositions, that is to say, it is possible to give a value
to all the properties in such a way of satisfying FUNC. This
possibility is lost in the quantum case and it is not a matter of
interpretation, it is the underlying mathematical structure that
enables this possibility for classical mechanics and forbids it in
the quantum case. In fact, one may also arrive at contextuality
from a topological analysis (Domenech and Freytes, 2005). To show
it, let us briefly recall that if  $I$ is a topological space, a
{\it sheaf} over $I$ is a pair $(A, p)$ where $A$ is a topological
space and $p:A \rightarrow I$ is a local homeomorphism. This means
that each $a\in A$ has an open set $G_a$ in $A$ that is mapped
homeomorphically by $p$ onto $p(G_a) = \{p(x): x\in G_a\}$, and
the latter is open in $I$. It is clear that $p$ is continuous and
open map.  {\it Local sections} of the sheaf $p$ are continuous
maps $\nu: U \rightarrow I$ defined over open proper subsets $U$
of $I$ such that the following diagram is commutative:

\begin{center}
\unitlength=1mm
\begin{picture}(60,20)(0,0)
\put(8,16){\vector(3,0){5}} \put(20,10){\vector(0,-2){5}}
\put(4,12){\vector(1,-1){10}}

\put(2,16){\makebox(0,0){$U$}} \put(20,16){\makebox(0,0){$A$}}
\put(20,0){\makebox(0,0){$U$}}

\put(2,20){\makebox(17,0){$\nu$}}
\put(6,10){\makebox(13,0){$\equiv$}}
\put(26,8){\makebox(-5,0){$p$}} \put(6,6){\makebox(-5,2){$1_U$}}
\end{picture}
\end{center}

\noindent In particular we use the term {\it global section} only
when $U=I$.

Thus, we may consider the family ${\cal W}$ of all Boolean
subalgebras of the lattice ${\cal L}({\mathcal H})$ ordered by
inclusion and the topological space $\langle {\cal W}, {\cal W}^+
\rangle$. On the set $E = \{(W,f): W\in {\cal W},\ f:W \rightarrow
{\bf 2}\ $ {\it is a Boolean homomorphism}$\}$ we define a partial
ordering given as $(W_1,f_1) \leq (W_2,f_2) \hspace{0.2cm}
\Longleftrightarrow \hspace{0.2cm} W_1 \subseteq W_2
\hspace{0.2cm} {\it and } \hspace{0.2cm} f_1 = f_2 \mid W_1$. Thus
we consider the topological space $\langle E, E^+ \rangle$ whose
canonical basis is given by the principal decreasing sets $((W,f)]
= \{(G ,f\mid G): G \subseteq W \}$. By simplicity $((W,f)]$ is
noted as $(W,f]$. Then, the map $p:E \rightarrow {\cal W}$ such
that $(W,f) \mapsto W$ is a sheaf over ${\cal W}$. We refer to it
as the {\it spectral sheaf}. We have shown elsewhere (Domenech and
Freytes, 2005) that, if $\nu: U \rightarrow E$ is a local section
of the spectral sheaf $p$, then for each $W\in U$ we have that
$\nu(W) = (W,f)$ for some Boolean homomorphism $f:W \rightarrow
{\bf 2}$ and also, if $W_0 \subseteq W$, then $\nu(W_0) = (W_0,
f\mid W_0)$. From the physical perspective, we may say that the
spectral sheaf takes into account the whole set of possible ways
of assigning truth values to the propositions associated with the
projectors of the spectral decomposition ${\bf A} = \sum_{i} a_{i}
{\bf P}_i$.

The continuity of a local section of $p$ guarantees that the truth
value of a proposition is maintained when considering the
inclusion of subalgebras. In this way, the {\it compatibility
condition} of the Boolean valuation with respect of intersection
of pairs of Boolean sublattices of ${\cal L}({\mathcal H})$ is
maintained. A global section $\tau: {\cal W} \rightarrow E $ of
$p$ is interpreted as follows: the map assigns to every $W \in
{\cal W}$ a fixed Boolean valuation $\tau_w:W \rightarrow {\bf 2}$
obviously satisfying the compatibility condition. So KS theorem in
terms of the spectral sheaf reads:

\begin{theo}\label{CS}
If $\mathcal{H}$ is a Hilbert space such that $dim({\cal H}) > 2$
then the spectral sheaf $p$ has no global sections.\qed \\
\end{theo}

\noindent We may build a {\it contextual valuation} in terms of a
local section as follows:  Let {\it A} be a physical magnitude
with known value, i.e. we have been able to establish a Boolean
valuation $f:W_A \rightarrow 2$. It is not very hard to see that
the assignment $\nu: (W] \rightarrow E$ such that for each $W_i\in
(W]$, $\nu(W_i) = (W_i, f\mid W_i)$ is a local section of $p$. To
extend contextual valuations we turn now to consider local
sections. To do this, let $\nu$ be a local section of $p$ and
$W_A$ the spectral algebra associated to the operator $\bf A$.
Then an extended valuation over $A$ is given by the set
$\bar{\nu}(A) = \{W_B \in dom(\nu) : W_B\subseteq W_A \}$ and it
is easy to prove that, if $\nu$ is a local section of $p$ and
$W_A$ the spectral algebra associated to the operator $\bf A$,
then: $\bar{\nu} (A)$ is a decreasing set and, if $W_A\in U$, then
$\bar{\nu}(A) = (W_A]$. Valuations are deeply connected to the
election of particular local sections of the spectral sheaf. So we
see here once more that we cannot speak of the value of a physical
magnitude without specifying this election, that clearly means the
election of a particular context.

\section{Contemporary Modal Interpretations of Quantum Mechanics}

Modal interpretations of quantum mechanics intend to provide an
objective reading of the mathematical formalism  ``in terms of
properties possessed by physical systems, independently of
consciousness and measurements (in the sense of human
interventions)'' (Dieks, 2005). Contemporary versions of the modal
approach were faced with the problem of doing so, staying away
from contradictions delivered by the KS theorem. B. van Fraassen
was the first one to formally include the reasoning of modal logic
in quantum mechanics. He presented a modal interpretation of
quantum logic in terms of its semantical analysis (van Fraassen
1973, 1981). This analysis had the purpose to clarify which
properties among those of the complete set structured in the
lattice of subspaces of Hilbert space pertain to the system:

{\smallroman
\begin{quotation}
``The interpretational question facing us is exactly: in general,
which value attributions are true? The response to this question
can be very conservative or very liberal. Both court later
puzzles. I take it that the Copenhagen interpretation -really, a
roughly correlated set of attitudes expressed by members of the
Copenhagen school, and not a precise interpretation- introduced
great conservatism in this respect. Copenhagen scientists appeared
to doubt or deny that observables even have values, unless their
state forces to say so. I shall accordingly refer to the following
very cautious answer as the {\it Copenhagen variant} of the modal
interpretation. It is the variant I prefer.'' B. van Fraassen
(1991, p. 280).
\end{quotation}}

After van Fraassen's interpretation, S. Kochen presented his own
modal version, also a continuation of the early founding fathers
discussions. C. F. von Weizs\"acker refers to it in the following
terms:

{\smallroman
\begin{quotation}
``We consider it is an illuminating clarification of the
mathematical structure of the theory, especially apt to describe
the measuring process. We would, however feel that it means not an
alternative but a continuation to the Copenhagen interpretation
(Bohr and, to some extent, Heisenberg).'' Th. G\"ornitz and C. F.
von Weizs\"acker (1987, p. 357)\end{quotation}}

In this section we intend to discuss the different versions of the
modal interpretation in order to delineate a general
characterization. We concentrate on van Fraassen's Copenhagen
variant, Kochen-Dieks modal interpretation, Bub's Bohmian variant
and  the atomic interpretation of Bacciagaluppi and Dickson,
because they comprise the main features of all modal
interpretations. (See also Healey 1989, Vermaas and Dieks 1995,
Bene and Dieks 2002; and Vermaas 1999a, Dickson 2002, Dieks and
Vermaas 1998 and de Ronde 2003 for general reviews) In particular,
we want to analyze the interpretational rules which account for
the path from the possible to the actual, and the meaning of
modality. For the sake of simplicity, we will be concerned here
only with pure states. This can be extended to density operators
as shown by Vermaas and Dieks (1995).

\subsection{Van Fraassen's Copenhagen Variant}

As noted by Dirac in the first chapter of his famous book, the
existence of superpositions is responsible of the striking
difference of quantum behaviour from classical one. In fact, the
photon being in a superposition of translational states must be
accepted if we want to explain interference effects (Dirac, 1958).
Superpositions are also central when dealing with the measurement
process, where the various terms associated with the possible
outcomes of a measurement must be assumed to be present together
in the description. This conduces to the so called ``measurement
problem''. Thus, the individual occurrence of a measurement
outcome is out of the scope of quantum mechanics,  and only the
probability of such result is given by the state of the system via
the Born rule. This fact leads van Fraassen to the distinction
between {\it value-attributing propositions} and {\it
state-attributing propositions}, between {\it value-states} and
{\it dynamic-states}:

{\smallroman
\begin{quotation}
``[...] a {\it state}, which is in the scope of quantum mechanics,
gives us only probabilities for actual occurrence of {\it events}
which are outside that scope. They can't be entirely outside the
scope, since the events are surely described if they are assigned
probabilities; but at least they are not the same things as the
states which assign the probability.

In other words, the state delimits what can and cannot occur, and
how likely it is -it delimits possibility, impossibility, and
probability of occurrence- but does not say what actually
occurs.'' B. van Fraassen (1991, p. 279)
\end{quotation}}

So van Fraassen distinguishes propositions about events and
propositions about states. Propositions about events are {\it
value-attributing propositions} $<{\bf A} , \sigma >$, they say
that `Observable ${\bf A}$ has certain value belonging to a set
$\sigma$'. Propositions about states are of the form `The system
is in a state of this or that type' (in a pure state, in some
mixture of pure states, in a state such that...). A {\it
state-attribution proposition} [${\bf A}, \sigma$] gives a
probability of the value-attribution proposition, it states that a
measurement of ${\bf A}$ must have an outcome in $\sigma$. {\it
Value-states} are specified by stating which observables have
values and which they are. {\it Dynamic-states}, by stating how
the system will develop if isolated, and how if acted upon in any
definite, given fashion. This is endowed with the following
interpretation:

{\smallroman
\begin{quotation}
``The interpretation says that, if a system $X$ has dynamic state
$\varphi$ at $t$, then the state-attributions [${\bf A}, \sigma$]
which are true are those that $Tr(\rho {\bf P}^{{\bf
A}}_{\sigma})=1$. [${\bf P}^{{\bf A}}_{\sigma}$ is the projector
over the corresponding subspace.] About the value-attributions, it
says that they cannot be deduced from the dynamic state, but are
constrained in three ways :
\begin{enumerate}
\item
If [${\bf A}, \sigma$] is true then so is the value-attribution
$<{\bf A}, \sigma>$: observable ${\bf A}$ has value in $\sigma$.
\item
All the true value-attributions could have Born probability 1
together.
\item
The set of true value-attributions is maximal with respect to the
feature (2.)'' B. van Fraassen (1991, p. 281).
\end{enumerate}
\end{quotation}}

This distinction between value-attribution propositions and
state-attribution propositions allows van Fraassen to face the
measurement problem from a new position. The way out of the
contradiction between the presence of various results associated
to the different terms in a superposition and the appearance of
only one result proposed by von Neumann is the projection
postulate, the non-causal state transition. In his spirit, an
observable pertaining to a system has a value if and only if the
system is in a corresponding eigenstate of the observable (the
eigenstate-eigenvalue link). So, the observable, say ${\bf A}$,
has a value if and only if a measurement of ${\bf A}$ is certain
to have certain outcome. If the outcome of the measurement is
uncertain, which is the case when the state is in a superposition
of eigenstates of the observable, then the observable has no
value.

Van Fraassen proposes to emphasize this modal character of the
theory via the r\^{o}le of the state:

{\smallroman
\begin{quotation}
``[...] the transition from the possible to the actual is not a
transition {\it of} state, but a transition {\it described by} the
state." B. van Fraassen (1991, p. 279)
\end{quotation}}

\noindent and to interpret the emergence of a result in a new
light:

{\smallroman
\begin{quotation}
``[...] [the emergence of a result is] {\it as if the Projection
Postulate were correct}. For at the end of a measurement of ${\bf
A}$ on system $X$, it is indeed true that  ${\bf A}$ has the
actual value which is the measurement outcome. But, of course, the
Projection Postulate is not really correct: there has been a
transition from possible to actual value, so what it entailed
about values of observables is correct, but that is all. There has
been no acausal state transition.'' B. van Fraassen (1991, p. 288)
\end{quotation}}

We now briefly recall the main aspects of van Fraassen's  modal
interpretation in terms of quantum logic, following (van Fraassen
1991, chap. 9). We concentrate on those aspects that will be
necessary in order to compare with our own treatment of modal
propositions (see section 5). In the modal interpretation, the
probabilities are of events, each describable as `an observable
having a certain value', which are aspects of the value states. If
$w$ is a physical situation in which system $X$ exists, then $X$
has both a {\it dynamic state} $\varphi$ and a {\it value state}
$\lambda$, i.e. $w=<\varphi, \lambda >$. A {\it value state}
$\lambda$ is a map of observable ${\bf A}$ into non-empty Borel
sets $\sigma$ such that it assigns \{1\} to $1_{\sigma}{\bf A}$.
$1_{\sigma}$ is the characteristic function of a set $\sigma$ of
values. So, if the observable $1_{\sigma}{\bf A}$ has value 1,
then it is impossible that ${\bf A}$ has a value outside $\sigma$.
The proposition $<{\bf A}, \sigma> = \{ w:\lambda (w)({\bf
A})\subseteq \sigma \}$ assigns values to physical magnitudes, it
is a {\it value-attribution proposition} and  is read as `${\bf
A}$ (actually) has value in $\sigma$'. $ {\cal V}$ is called the
set of value attributions ${\cal V}=\{ <{\bf A}, \sigma>:{\bf A}\
{\rm an \ observable \ and}\ \sigma \ {\rm a\ Borel\ set}\}$. The
logic operations among value-attribution propositions are defined
as: $<{\bf A}, \sigma>^{\bot} = <{\bf A}, \Re - \sigma>$, $<{\bf
A}, \sigma>\wedge <{\bf A}, \theta>= <{\bf A}, \sigma\cap
\theta>$, $<{\bf A}, \sigma>\vee <{\bf A}, \theta>= <{\bf A},
\sigma\cup \theta>$ and $\wedge\{ <{\bf A}, \sigma_{i}
>: i \in {\cal N}\} = <{\bf A}, \cap\{ \sigma_{i}: i \in {\cal
N}\}>$. With all this, ${\cal V}$ is the union of a family of
Boolean sigma algebras $<{\bf A}>$ with common unit and zero equal
to $<{\bf A}, S({\bf A})>$ and $<{\bf A}, \wedge >$ respectively.
The Law of Excluded Middle is satisfied: every situation $w$
belongs to $q\vee q^{\bot}$, but not the Law of Bivalence:
situation $w$ may belong neither to $q$ nor to $q^{\bot}$.

A {\it dynamic state} $\varphi$ is a function from ${\cal V}$ into
[0, 1], whose restriction to each Boolean sigma algebra $<{\bf
A}>$ is a probability measure. The relation between dynamic and
value state is the following: $\varphi$ and $\lambda$ are a
dynamic state and a value state respectively, only if there exist
possible situations $w$ and $w'$ such that $\varphi =\varphi (w),\
\lambda=\lambda (w')$. Here, $\varphi$ is an eigenstate of ${\bf
A}$, with corresponding eigenvalue $a$, exactly if $\varphi(<{\bf
A}, \{ a\}>)=1$. The {\it state-attribution proposition} [${\bf
A},\sigma$] is defined as: [${\bf A}, \sigma$] = $\{ w:
\varphi(w)(<{\bf A}, \sigma>) =1\}$ and means `${\bf A}$ must have
value in $\sigma$'. ${\cal P}$ denotes the set of
state-attribution propositions: ${\cal P}=\{ [{\bf A}, \sigma]:
{\bf A} \ {\rm an \ observable, }\ \sigma \ {\rm a \ Borel \
set}\}$. Partial order between them is given by $[{\bf A}, \sigma
]\subseteq [{\bf A}^{'} , \sigma^{'}]\ {\rm only\ if,\ for\ all\
dynamical\ states} \ \varphi, \ \varphi(<{\bf A}, \sigma>) \leq
\varphi(<{\bf A}^{'} , \sigma^{'} >)$ and the logic operations are
(well) defined as: $[{\bf A}, \sigma]^{\bot} = [{\bf A}, \Re
-\sigma ]$, $[{\bf A}, \sigma]\uplus [{\bf A}, \theta]= [{\bf A},
\sigma\cup \theta ]$ and $[{\bf A}, \sigma] \cap [{\bf A}, \theta]
= [{\bf A}, \sigma\cap \theta]$. With all this, $<{\cal P},
\subseteq , ^{\bot}>$ is an orthoposet, the orthoposet formed by
`pasting together' a family of Boolean algebras in which whole
operations coincide in areas of overlap. It may be enriched to
approach the lattice of subspaces of  Hilbert space.

One may recognize a modal relation between both kind of
propositions. One starts denying the collapse in the measurement
process  and recognizing that the observable has one of the
possible eigenvalues. Then it may be asked what may be inferred
with respect of those values when one knows the dynamic state. The
answer van Fraassen gives is that, in the case that $\varphi(w)$
is an eigenstate of the observable ${\bf A}$ with eigenvalue $a$,
then ${\bf A}$ actually does have value $a$. This means that in
this case, the measurement `reveals' the value the observable
already had. He generalizes this idea and  postulates that [${\bf
A}, \sigma$] implies $<{\bf A}, \sigma>$. With this assumption and
the rejection of an ignorance interpretation of the uncertainty
principle, he is able to prove that [${\bf A}, \sigma$] $= \Box
<{\bf A}, \sigma>$. $\Box$ is defined by $\Box Q = \{ w:\ {\rm
for\ all\ }w',\ {\rm if}\ wRw'\ {\rm then}\ w' \in Q\} $, where
$q$ is any proposition and $R$ the relative possibility relation:
$w'$ is possible relative to $w$ exactly if, for all $Q$ in ${\cal
V}$, if $w$ is in $Q$ then $w'$ is in $Q$. So, [${\bf A}, \sigma$]
may be read as `necessarily, $<{\bf A}, \sigma>$'. This says that
the dynamic state assigns 1 to $<{\bf A}, \sigma>$ if and only if
the value state that accompanies any relatively possible dynamic
state makes $<{\bf A}, \sigma>$ true. Instead of the transitive
possibility relation $R$, one may use an equivalence relation to
define the necessity operator $\widetilde{\Box}$. In this case,
van Fraassen maintains that the map [${\bf A}, \sigma$]
$\rightarrow <{\bf A}, \sigma>$ is an isomorphism of posets
$<{\cal P}, \subseteq>$ and $<{\cal V}, \subseteq>$ and, when
orthocomplementation if defined, it becomes an isomorphism between
the orthoposets. Thus, the logic of ${\cal V}$ is that of ${\cal
P}$, i.e., quantum logic. Endowed with these tools, van Fraassen
gives an interpretation of the probabilities of the measurement
outcomes which is in agreement with the Born rule. To do this, he
considers mixtures. However, in spite of his motivation, we will
continue restricting to the case of pure states because our aim is
to analyze structures which gather together possibility and
actuality. In fact, our approach allows us to organize a
propositional system that takes into account the `paste' of
propositions about actual properties and modal properties.
(Domenech {\it et al.}, 2006a)

\subsection{Kochen-Dieks Modal Interpretation}

The next modal interpretation we would like to review is due to S.
Kochen and D. Dieks (K-D, for short), who proposed to use the so
called biorthogonal decomposition theorem (also called Schmidt
theorem) in order to describe the correlations between the quantum
system and the apparatus in the measurement process:

\begin{theo}\label{Schmidt}
Given a state $|\Psi_{\alpha\beta}\rangle$ in $\cal H = \cal
H_{\alpha}\otimes \cal H_{\beta}$. The Schmidt theorem assures
there always exist orthonormal bases for $\cal H_{\alpha}$ and
$\cal H_{\beta}$, $\{|a_{i}\rangle\}$ and $\{|b_{j}\rangle\}$ such
that $|\Psi_{\alpha\beta}\rangle$ can be written as:

\begin{center}$|\Psi_{\alpha\beta}\rangle = \sum c_{j}|a_{j}\rangle\
\otimes |b_{j}\rangle$.\end{center}

\noindent The different values in $\{|c_{j}|^{2}\}$ represent a
spectrum of the Schmidt decomposition given by $\{\lambda_{j}\}$.
Every $\lambda_{j}$ represents a projection in $\cal H_{\alpha}$
and a projection in $\cal H_{\beta}$ defined as
$P_{\alpha}(\lambda_{j}) = \sum |a_{j}\rangle\ \langle a_{j}|$ and
$P_{\beta}(\lambda_{j}) = \sum |b_{j}\rangle\ \langle b_{j}|$,
respectively. Furthermore, if the $\{|c_{j}|^{2}\}$ are non
degenerate, there is a one-to-one correlation between the
projections $P_{\beta} (\lambda_{j})$ and
$P_{\alpha}(\lambda_{j})$ pertaining to subsystems $\cal
H_{\alpha}$ and $\cal H_{\beta}$ given by each value of the
spectrum $\lambda_{j}$.\qed
\end{theo}

\noindent Through this theorem one is able to distinguish, by
tracing over the degrees of freedom of the subspace $\cal
H_{\alpha}$ or the subspace $\cal H_{\beta}$, between system and
apparatus (for a proof of the theorem see Bacciagaluppi 1996,
section 2.3). As noted by Kochen (1985, p. 152): ``Every
interaction gives rise to a unique correlation between certain
canonically defined properties of the two interacting systems.
These properties form a Boolean algebra and so obey the laws of
classical logic." The biorthogonal decomposition gives in this
way, a one to one relation between the apparatus and the quantum
system and the following interpretation: {\it The system $\alpha$
possibly possesses one of the properties $\{|a_{j}\rangle\ \langle
a_{j}|\}$, and the actual possessed property $|a_{k}\rangle\
\langle a_{k}|$ is determined by the observation that the device
possesses the reading $|b_{k}\rangle\ \langle b_{k}|$.} However,
by tracing over the degrees of freedom of the system, one obtains
an \emph{imporper mixture}. It is well known that improper
mixtures cannot be interpreted in terms of ignorance (D'Espagnat,
1976), and thus, one comes back to the problem of interpreting
modalities. Following van Fraassen's distinction between value
states and dynamical states, Dieks solves the problem of putting
together the seemingly incompatible character of improper mixtures
and ignorance via the distinction between different levels of
discourse. For example, with respect to Schr\"odinger's cat, Dieks
(1988a, p. 189) states that: ``It is the state vector which is in
a superposition, not the cat itself. `State vector' and `cat' are
two concepts at different levels of discourse.'' In order to
explicitly take into account  this remark, P. Vermaas and D. Dieks
distinguish between {\it physical properties} and {\it
mathematical properties}:

{\smallroman
\begin{quotation}
``[...] the interpretation gives the probabilities of various
possible states of affairs of which it is assumed that only one is
actually realized. [...]  it is crucial to notice the distinction
which is made in the modal interpretation between physical
properties and mathematical states. The latter are defined in
Hilbert space and encode probabilistic information about the
properties (values of physical magnitudes) which are present in
the system itself. There is no one-to-one relation between
physical properties and mathematical states. Now, the ignorance
inherent in the modal interpretation pertains to physical
properties; not to mathematical states.'' P. Veramaas and D. Dieks
(1995, p. 155)
\end{quotation}}

Once again, the main problem remains that of interpreting
possibility and its relation to actuality:

{\smallroman
\begin{quotation}
``[...] an irreducible statistical theory only speaks about
possible outcomes, not about the actual one; this predicts only
probability distributions of all outcomes, and says nothing about
the result which really will be realized in a single case. In
brief, such a theory is not about what is real and actual but only
about what could be the case.'' D. Dieks (1988a, p. 177)
\end{quotation}}

\noindent Kochen (1985, p. 152) and Dieks do not consider the
collapse of the state vector as a physical process: ``[...] the
collapse of the wave function is not a real physical effect in
this interpretation but is simply the result of a change in
perspective from our witnessing system to another.'' However, they
must give an account of the emergence of a single result, they
thus appeal to an interpretational rule in an analogous fashion to
van Fraassen's proposal:

{\smallroman
\begin{quotation}
``I now propose the following interpretational rule: as soon as
there is a unique decomposition of the form (2), the partial
system represented by the $|\phi_{k}\rangle$, taken by itself, can
be \emph{described} as possessing one of the values of the
physical quantity corresponding to the set ${|\phi_{k}\rangle}$,
with probability $|c_{k}|^{2}$. [By Eq (2), they mean
$|\Psi_{\alpha\beta}\rangle = \sum c_{j}|\phi_{j}\rangle\ \otimes
|R_{j}\rangle$]

This rule is intended to have the following important consequence.
Experimental data that pertain only to the object system, and that
say it possesses the property associated with, e.g.,
$|\phi_{1}\rangle$, not only count as support for the theoretical
description $|\phi_{1}\rangle|R_{1}\rangle$ but also as empirical
support for the theoretical description (2)." D. Dieks (1988b, p.
39)
\end{quotation}}

It is important to notice that the seemingly \emph{ad hoc} move of
using a preferred basis (such as the Schmidt basis) can be given a
physical motivation. It has been proved by Dieks (1995) that,
given the following two conditions:

\begin{enumerate}
\item {one-to-one correlation}: we require a one to
one correlation between the definite properties of the system and
the definite properties of its environment,

\item {no hidden variables}: the Hilbert space formalism,
with the usual representation of physical magnitudes by
observables, should be completely respected,
\end{enumerate}

\noindent the only basis that accomplishes these two conditions is
the Schmidt basis. The first demand appears as obvious when
reflecting on the preconditions which allow us to talk about
measurement. The second demand can be considered as a commitment
to the early interpretation of Bohr, Born, Heisenberg and Pauli;
to consider the quantum description as providing all there is to
know with respect to atomic events, and that one should not seek
for a more complete causal description in terms of space and time.
As expressed by Bohr (1928), and also by Heisenberg (1949, section
3), these two descriptions are complementary to each other (see
also Dieks, 1989, section 9).

It is important to stress at this point that, given the complete
system and its corresponding Hilbert space,  the choice of what is
the system under study and what is the apparatus determines the
factorization of the complete space that leads to the set of
definite properties given by the Schmidt decomposition. In spite
of the fact that this cut is (mathematically) not fixed {\it a
priori} in the formalism, the (physical) choice of the apparatus
determines explicitly the context. It is exactly this possibility,
of having different incompatible contexts given by the choice of
mutually incompatible apparatuses, which in turn determines KS
type contradictions within the modal interpretation (see, for a
discussion, Karakostas 2004, section 6.1).

As noted by Dieks in a recent paper (Dieks, 2007): ``According to
the modal interpretation the state in Hilbert space thus is about
possibilities, about what may be the case; about modalities. But
there is also a second aspect to $\psi$: it is the theoretical
quantity that occurs in the evolution equation, and its evolution
governs deterministically how the set of definite valued
quantities changes. This double role of $\psi$, on the one hand
probabilistic and on the other dynamical and deterministic, is a
well-known feature of the Bohm interpretation." We will now turn
our attention to Bub's Bohmian variant in order to discuss further
the relation between Bohmian mechanics and the modal
interpretation.

\subsection{Bub's Bohmian Variant}

The modal version of J. Bub recalls on D. Bohm's interpretation
and proposes to take some observable, $\bf R$, as {\it always}
possessing a definite value. In this way one can avoid KS
contradictions and maintain a consistent discourse about
statements which pertain to the sub-lattice determined by the {\it
preferred observable} $\bf R$. As van Fraassen's and Vermaas and
Dieks' interpretations, Bub's proposal distinguishes between {\it
dynamical states} and {\it property} or {\it value states}, in his
case with the purpose of interpreting the wave function as
defining a Kolmogorovian probability measure over a restricted
subalgebra of the lattice ${\cal L}({\mathcal H})$ of projection
operations (corresponding to yes-no experiments) over the state
space. It is this distinction between property states and
dynamical states that gives a modal character to the
interpretation:

{\smallroman
\begin{quotation}
``The idea behind a `modal' interpretation of quantum mechanics is
that quantum states, unlike classical states, constrain
possibilities rather than actualities -which leaves open the
question of whether one can introduce property states [...] that
attribute values to (some) observables of the theory, or
equivalently, truth values to the corresponding propositions.'' J.
Bub (1997, p. 173)
\end{quotation}}

In precise terms, as ${\cal L}({\mathcal H})$ does not admit a
global family of compatible valuations, and thus not all
propositions about the system are determinately true or false,
probabilities defined by the (pure) state cannot be interpreted
epistemically (Bub, 1997, 119). But, if one chooses, for a given
state $|e>$, a ``preferred observable'' $\bf R$, these properties
can be taken as determinate since the propositions associated with
$\bf R$, i.e., with the projectors in which $\bf R$ decomposes,
generate a Boolean algebra. Bub constructs the maximal sublattices
${\cal D}(|e>, \ {\bf R}) \subseteq {\cal L}({\cal H})$ to which
truth values can be assigned via a 2-valued homomorphism and
demonstrates a uniqueness theorem that allows the construction of
the preferred observable.

In Bub's proposal, a {\it property state} is a maximal
specification of the properties of the system at a particular
time, defined by a Boolean homomorphism from the determined
sublattice to the Boolean algebra of two elements. On the other
hand, a {\it dynamical state} is an atom of ${\cal L}({\cal H})$
that evolves unitarily in time following Schr\"{o}dinger equation.
So, dynamical states do not coincide with property states. Given a
dynamical state represented by the atom $|e>\in{\cal L}({\cal
H})$, one constructs the sublattice ${\cal D}(|e>, \ {\bf R})$
with Kolmogorovian probabilities defined over alternative subsets
of properties in the sublattice. They are the properties of the
system, and the probabilities defined by $|e>$ evolve (via the
evolution of $|e>$) in time. If the preferred observable is the
identity operator ${\bf I}$, the atoms in ${\cal D}(|e>, \ {\bf
I})$ may be pictured as a ``fan'' of its projectors generated by
the ``handle'' $|e>$ (Bub, 1992, 751) or an ``umbrella'' with
state $|e>$ again as the handle and the rays in $(|e>)^{\bot}$ as
the spines. When observable ${\bf R}\neq{\bf I}$, there is a set
of handles $\{ |e_{r_{i}} >, i=1...k\}$ given by the nonzero
projections of $|e>$ onto the eigenspaces of ${\bf R}$ and the
spines represented by all the rays in the orthogonal complement of
the subspace generated by the handles. When $dim({\cal H})>2$,
there are $k$ 2-valued homomorphisms which map each of the handles
onto 1 and the remaining atoms onto 0. The determinante sublattice
(that changes with the dynamics of the system) is a partial
Boolean algebra, i.e., the union of a family of Boolean algebras
pasted together in such a way that the maximum and minimum
elements of each one, and eventually other elements, are
identified and, for every $n$-uple of pair-wise compatible
elements, there exists a Boolean algebra in the family containing
the $n$ elements. The possibility of constructing a probability
space with respect to which the Born probabilities generated by
$|e>$ can be thought as measures over subsets of property states
depends on the existence of sufficiently many property states
defined as 2-valued homomorphisms over ${\cal D}(|e>,\ {\bf R})$.
This is guaranteed by a uniqueness theorem that characterizes
${\cal D}(|e>,\ {\bf R})$ (Bub, 1997, 126). Thus constructed, the
structure avoids KS-type theorems. Then, given a system $S$ and a
measuring apparatus $M$,

{\smallroman
\begin{quotation}
``[...] if some quantity $\bf R$ of $M$ is designated as always
determinate, and $M$ interacts with $S$ via an interaction that
sets up a correlation between the values of $\bf R$ and the values
of some quantity $\bf A$ of $S$, then $\bf A$ becomes determinate
in the interaction. Moreover, the quantum state can be interpreted
as assigning probabilities to the different possible ways in which
the set of determinate quantities can have values, where {\it one
particular set of values represents the actual but unknown values
of these quantities}.'' J. Bub (1992, p. 750, emphasis added)
\end{quotation}}

The problem with this interpretation is that, in the case of an
isolated system,  there is no single element in the formalism of
quantum mechanics which allows us to choose an observable, ${\bf
R}$, rather than other.  This is why the move seems fragrantly
{\it ad hoc}. Of course, were we dealing with an apparatus, there
would be a preferred observable, namely the pointer position.

Our research is focused exactly on the meaning of modality. It is
thus interesting to go back to Bohm himself and recover the ideas
which allowed him to develop his ``causal interpretation". As
noted by Bub with respect to Bohmian mechanics:

{\smallroman
\begin{quotation}
``[...] the change in the quantum state $|\psi\rangle$ manifests
itself directly at a {\it modal} level --the level of possibility
rather than actuality-- through the determinate sublattice defined
by $|\psi\rangle$ and position in configuration space as the
preferred determinate observable.'' J. Bub (1997, p. 170)
\end{quotation}}

In (Bohm, 1953)  Bohm proves that the probability density in his
interpretation approaches the quantum probability density
$|\psi|^{2}$. In this paper Bohm clearly recognizes the main
distinction between the usual interpretation, provided by Bohr,
Born, Heisenberg and Pauli, and his own causal interpretation:

{\smallroman
\begin{quotation}
``The arbitrariness of the usual interpretation in the description
of the behavior of an individual system is closely related to the
assumption, already stated, that the wave function determines all
physically significant properties of the system. Now, in the case
of the uranium nucleus, the wave function takes the form of a
packet initially entirely within the nucleus, which gradually
`leaks' through the barrier and thereafter rapidly spreads without
limit in all directions. Clearly, although this wave function is
supposed to describe \emph{all} physically significant properties
of the system, it cannot explain the fact that each
$\alpha$-particle is actually detected in a comparatively small
region of space and at a fairly well-defined instant of time.
\emph{The usual interpretation states that this phenomenon must be
simply accepted as an event that somehow manages to occur but in a
way that is as a matter of principle forever beyond the
possibility of a simultaneous and detailed `space-time and causal
description.'} Indeed, even to ask for such a description is said
to be a meaningless question within the framework of the usual
interpretation of the quantum theory. \emph{In the causal
interpretation, however, the postulated particles with precisely
defined positions explain in a natural way why an
$\alpha$-particle can be detected as a fairly definite place and
time, on the basis of the simple assumption that the particle
existed all the time and just moved from its original location to
the place where it was finally found.} Thus, even though we cannot
yet observe the precise locations of our postulated particles,
they already perform a real function in the theory, namely, to
explain certain properties of \emph{individual} systems which are
said in the usual interpretation to be just empirically given and
forever unexplainable." D. Bohm (1953, p. 464, emphasis added)
\end{quotation}}

\noindent The usual interpretation follows the second demand given
also by Dieks (1995), that no deterministic picture can provide a
more complete description of the state of affairs. Bohm's causal
interpretation, on the other hand, wishes to retain this classical
feature.

{\smallroman
\begin{quotation}
``[...] \emph{in the usual interpretation two completely different
kinds of statistics are needed.} First, there is the ordinary
statistical mechanics, which treats of the distortion of systems
among the quantum states, resulting from various chaotic factors
such as collisions. The need of this type of statistics could in
principle be avoided by means of more accurate measurements which
would supply more detailed information about the quantum state,
but in systems of appreciable complexity, such measurements would
be impracticably difficult. Secondly, however, there is the
fundamental and irreducible probability distribution,
$P(x)=|\psi(x)|^{2}$ [...]. The need of this type of statistics
cannot even in principle be avoided by means of better
measurements, nor can it be explained in terms of the effects of
random collision processes. [...] On the other hand, \emph{the
causal interpretation requires only one kind of probability.} For
as we have seen, we can deduce the probability distribution
$P(x)=|\psi(x)|^{2}$ as a consequence of the same random collision
processes that give rise to the statistical distributions among
the quantum states." D. Bohm (1953, p. 465, emphasis added)
\end{quotation}}

\noindent These two statistics to which Bohm refers are the two
levels of discourse present in modal interpretations (dynamical
and value state). Bohm wishes to recover the classical concept of
probability as lack of knowledge, however, as it will become clear
in section 5, this is not achievable, in general, in the modal
scheme. Possibility remains a contextual concept: a set of shared
possibilities corresponding to disjoint sets of actual properties
cannot be non-contextually actualized without contradictions.
Classical probability can only be recovered as lack of knowledge
once  the definite context has been chosen.

\subsection{Atomic Modal Interpretation}

The atomic modal interpretation is due to G. Bacciagaluppi and M.
Dickson (1997). It intends, via a factorization, to separate the
state space of the system $\cal H$ in disjoint spaces ${\cal
H}_{k}$. A factorization $\Phi$ of a Hilbert space $\cal H$ into a
tensor product of two Hilbert spaces ${\cal H}_{1}\otimes {\cal
H}_{2}$ is given by an equivalence class of isomorphisms differing
only by a basis transformation of the factor spaces onto
themselves. It may be proved that there are many different
factorizations. The question becomes now whether, by letting
$\Phi$ vary, the definite properties pertaining to the different
factorizations will admit a truth valuation. Bacciagaluppi has
proved that this question must be answered negatively because
these properties include the set of properties for which KS have
shown that it is not allowed an homomorphism to the Boolean
algebra {\bf 2} (Bacciagaluppi, 1995). In order to escape this
no-go theorem, Bacciagaluppi and Dickson assume that there exists
in nature a special set of disjoint spaces ${\cal H}_{k}$ which
are the building blocks of all physical systems; i.e. a {\it
preferred factorization}  of the Hilbert space:

{\smallroman
\begin{quotation}
``[...] we note that the idea of a preferred factorization is not,
perhaps, as {\it ad hoc} as it might first appear. After all
assuming that the universe is really made of, say, electrons,
quarks, and so on, it makes good sense to take these objects to be
`real' constituents of the universe, i.e. the bearers of
properties that do not supervene on the properties of subsystems."
G. Bacciagaluppi and M. Dickson (1997, p. 3)\end{quotation}}

It is important to notice that in such interpretation the
structure of the probability assignment becomes classical, i.e.
one can define a classical joint probability distribution for any
set of chosen properties. As a consequence, probability can be
interpreted in terms of ignorance. In the atomic interpretation
there is a single context given by the preferred factorization and
thus, as in classical physics,  the KS theorem does not apply. But
the intrinsic modal aspect we have been referring up to now does
not have to do with ignorance and, in this sense, the atomic modal
interpretation may be considered closer to classical statistical
interpretations of quantum mechanics (see, for example,
Ballentine, 1990).

Bacciagaluppi's interpretation of modal interpretations
(Bacciagaluppi, 1996), appears in an analogous fashion to Bub's
proposal, closely related to Bohm's causal interpretation.

{\smallroman
\begin{quotation}
``The properties possessed by a system in the modal interpretation
are possessed {\it in addition} to the properties possessed by the
system according to quantum mechanics. It is thus natural to call
these properties `hidden variables'. Hidden variables theories do
not represent a return to classical, pre-quantum physics. Indeed,
the no-go theorems for hidden variables theories show not that
hidden variables are impossible, but that they must be in
important ways different from classical physics (e.g. they are
non-local). On the other hand, hidden variables theories always
restore a classical way of thinking about \emph{what there is.} In
particular, the logical and probabilistic structure of a hidden
variables theory is always classical: there is no
`complementarity' of hidden variables, and probabilities are
rigorously Kolmogorovian. [...] The status of probabilities in the
modal interpretation, however, has been the subject of some
debate, I would presume partially because the original modal
interpretation of van Fraassen was not intended as a theory about
what there is, but, indeed, as a theory about possibilities. Thus,
Dickson (1995b) has wondered whether the modal interpretation
really is a no-collapse interpretation, and Healey (1995) has
expressed reservations about the desirability of introducing a
dynamics for the proposed properties. However, for our proposes, I
would claim that, despite the name, the modal interpretation in
the version of Vermaas and Dieks is a theory about actualities --
albeit a stochastic one." G. Bacciagaluppi (1996, p.
74)\end{quotation}}

\noindent Bacciagaluppi is correct to point out that if one takes
properties in the modal interpretation as existing in actuality,
one could argue that modal interpretations are in the end, some
kind of hidden variable theory. However, Bacciagaluppi does not
recognize that the distinction between dynamical state and value
state in van Fraassen's interpretation, or mathematical and
physical state in Vermaas Dieks proposal, provides exactly this
non-Kolmogorovian model regarding properties, and thus, the `two
statistics' needed in the usual interpretation, and criticized by
Bohm. The mode of existence regarding the properties in the
dynamical state (or in the mathematical state) is what provides a
formal picture which remains non-classical, and which cannot be
interpreted in terms of actuality but only as possibility, or
maybe even in terms of potentiality (de Ronde, 2005, section 1.4).
Bacciagaluppi and Dickson regard modal interpretations as
referring to actual properties, this is why they look for a
dynamical picture that governs the evolution of these properties.
On the contrary, this is considered by van Fraasen and Dieks as
superfluous.

\section{A General Characterization of Modal Interpretations of
Quantum Mechanics}

Taking into account our previous analysis we are now in conditions
to summarize what we consider the general characteristics of modal
interpretations.

\begin{enumerate}
\item
One of the most significant features of modal interpretations is
that they stay close to the standard formulation. Following van
Fraassen's recommendation, one needs to learn from the formal
structure of the theory in order to develop an interpretation.
This is different from many attempts which presuppose an ontology
and then try to fit it into the formalism.

\item
Modal interpretations are non-collapse interpretations. The
evolution is always given by the Schr\"odinger equation of motion
and the collapse of the wave function is nothing but the path from
the possible to the actual, it is not considered a physical
process.

\item
Modal interpretations ascribe possible properties to quantum
systems. The property ascription depends on the states of the
systems and applies regardless of whether or not measurements are
performed.  There is a distinction between the level of
possibility and that of actuality which are related through an
interpretational rule. In addition to actual properties
interpreted in the orthomodular lattice of subspaces of ${\cal
H}$, there is a set of possible properties which may be regarded
as  constituting the center of the enriched lattice (see section
5).

\item
Modality is not interpreted in terms of ignorance. There is no
ignorance interpretation of the probability distribution assigned
to the physical properties. The state of the system determines all
there is to know. For modal interpretations there is no such thing
as `hidden variables' from which we could get more information.
One can formulate a KS theorem for modalities which expresses the
irreducible contextual character of quantum mechanics even in the
case of enriching its language with a modal operator (see section
5).

\end{enumerate}

It is easily seen that Born's and specially Heiseneberg's
interpretations present already the basic features of modal
interpretations; these ideas would be later rigourously formalized
by  van Fraassen's and Kochen-Dieks' interpretations. Van
Fraassen's account is {\it modal} because it leads, in a
relatively straightforward way, to a modal logic of quantum
propositions. In the Copenhagen variant, the most important point
is perhaps that one should not presume this modal logic to arise
from ignorance about the actual state of affairs, which is the aim
of science to uncover. In other words, we do not say that a system
with dynamical state $\varphi$ possibly has some value state and
we need to find out which one, or which one with which
probability. What is important is that there are possible value
states for all physical systems (i.e. possible stories about the
world) that are compatible with all the observable data. Van
Fraassen is closer to QL and distinguishes two kind of
propositions related by a modal quantifier. However, he defines
himself as an empiricist and does not interpret modality in an
ontological fashion, remaining agnostic about the relation between
modalities and `the world'. K-D present a different property
ascription rule given by the Schmidt decomposition. They are able
to ascribe a set of definite properties to the system giving, in
this way, a realistic flavor to the interpretation (see Dickson,
2002). For Dieks, one of the most pleasant features of modal
interpretations is that they remain within the orthodox
formulation of quantum mechanics, there is no need of hidden
variables and the state of the system gives all the information
there is to know. Maybe the most characteristic feature of both
van Fraassen and Dieks interpretations, which go back to
Heisenberg and Bohr, is the distinction between different levels
of discourse; i.e. dynamical and value states in van Fraasen, and
the mathematical and physical states, in Vermaas and Dieks.

The main ideas which guide the line of investigation of the
Bohmian causal interpretation go against points 1, 3 and 4 of our
characterization of modal interpretations. In both atomic and
Bohmian variant, ``possibility" is taken to express a degree of
ignorance of a determined state of affairs. In Bub's proposal,
ignorance about the exact \emph{position} of the particles while,
in the atomic interpretation, ignorance regarding which is the
\emph{preferred factorization} of the Universe. In both
interpretations the state of system is not all there is to know,
possibility is only regarded in terms of actuality and, as noticed
by Bacciagaluppi (1996, p. 74), modal interpretations thus become
``a theory about actualities". This is why in both atomic and
Bub's interpretations, there is no distinction between different
levels of discourse. That which exists, exists only at the level
of actuality, exactly in the same sense as in classical
statistical mechanics. In spite of the fact that one may state
that this is the same empirical statement of van Fraassen, one
must recall that in van Fraassen's interpretation, even though
that which exists is actual (van Fraassen, 1981, section 5.3),
there is a level of possibility given by the dynamical state of
which he remains agnostic but which has nothing to do with
ignorance.

Summarizing, on the one hand, in van Fraassen Copenhagen variant
and K-D modal interpretation it remains clear that modality cannot
be interpreted in terms of ignorance. However, in both
interpretations the ontological significance of possibility
remains untouched. The idea which van Fraassen sustains is that:
{\it modalities are in our theories, not in the world}. Contrary
to this position, one of us has proposed to interpret possibility
in terms of ontological potentiality (de Ronde, 2005), taking to
its last consequences the ideas of Heisenberg, and presenting
possibility in terms of Aristotle, as another mode of existence,
complementary to that of actuality. On the other hand, in Bub's
Bohmian interpretation and also in the atomic one, modality is
still thought in terms of ignorance, remaining closer to a
classical statistical conception of modality. This is the
character which we are specifically interested in discussing in
detail. In the following section we will analyze, taking into
account previous work (Domenech {\it et al.}, 2006a; Domenech {\it
et al.}, 2006b) whether this idea can be consistently maintained
in the orthodox formalism of quantum mechanics.

\section{The Contextual Character of Modal Interpretations of Quantum Mechanics}

In order to stay away from inconsistencies when speaking about
properties which pertain to the system, one must acknowledge the
limitations imposed by the KS theorem. To do so, modal
interpretations assign to the system only a set of definite
properties. However, it has been shown by Bacciagaluppi (1995), R.
Clifton (1995, 1996) and later by Vermaas (1997, 1999b) that this
is not achievable when talking about properties which pertain to
different contexts (see also Bacciagaluppi and Vermaas, 1999).

At first sight it might seem paradoxical that, even though quantum
mechanics talks about modalities, KS theorem refers to actual
values of physical properties. Elsewhere, and following the line
of thought of quantum logic, we have investigated the question
whether KS theorem has something to say about possibility and its
relation to actuality (Domenech {\it et al.}, 2006a; Domenech {\it
et al.}, 2006b). The answer was provided via a characterization of
the relations between actual and possible properties pertaining to
different contexts. By applying algebraic and topological tools we
studied the structure of the orthomodular lattice of actual
propositions enriched with modal propositions. Let us briefly
recall the results.  As usual, given a proposition about the
system, it is possible to define a context from which one can
predicate with certainty about it (and about a set of propositions
that are compatible with it) and predicate probabilities about the
other ones. This is to say that one may predicate truth or falsity
of all possibilities at the same time, i.e. possibilities allow an
interpretation in a Boolean algebra. In rigorous terms, for each
proposition $P$, if we refer with $\Diamond P$ to the possibility
of $P$, then $\Diamond P$ will be a central element, i.e., a
complemented element which satisfies the distributive law for
every set of three elements of the lattice. The orthomodular
lattice thus expanded includes propositions about possibility. If
$P$ is a proposition about the system and $P$ occurs, then it is
trivially possible that $P$ occurs. This is expressed as $P \leq
\Diamond P$. If we identify $P$ with the value-attribution
proposition $<{\bf A}, \sigma>$ as defined by van Fraassen, we may
say that the classical consequences of $P$ coincide with those of
its correspondent state-attribution proposition [${\bf A},
\sigma$]. In fact, to assume an actual property and a complete set
of properties that are compatible with it determines a context in
which the classical discourse holds. Classical consequences that
are compatible with it, for example probability assignments to the
actuality of other propositions, shear the classical frame. These
consequences are the same ones as those which would be obtained by
considering the original actual property as a possible property.
This is interpreted as, if $P$ is a property of the system,
$\Diamond P$ is the smallest central element greater than $P$.
With these tools, we are able to give an extension of the
orthomodular structure by adding a possibility operator that
fulfills the mentioned requirements. More precisely, the extension
is a class of algebras, called Boolean saturated orthomodular
lattices, that admits the orthomodular structure as a reduct and
we demonstrate that they are a variety, i.e., definable by
equations. Complete orthomodular lattices are examples of them
(Domenech {\it et al}, 2006a). We have also found its logic
(Domenech {\it et al.}, 2007b). This algebraic construction is
what allows us to consistently expand the structure of actual
properties to include modal properties. This is so because we have
proved that every orthomodular lattice may be embedded in a
Boolean saturated orthomodular one.

If ${\cal L}$ is an orthomodular lattice and ${\cal L}^{\Diamond}$
a Boolean saturated orthomodular one such that ${\cal L}$ can be
embedded in ${\cal L}^{\Diamond}$, we say that ${\cal
L}^{\Diamond}$ is a modal extension of ${\cal L}$. Given  ${\cal
L}$ and a modal extension ${\cal L}^{\Diamond}$,  we define the
{\it possibility space} as the subalgebra of ${\cal L}^{\Diamond}$
generated by $ \{\Diamond P : P \in {\cal L} \} $. We denote by
$\Diamond {\cal L}$ this space and it may be proved that it is a
Boolean subalgebra of the modal extension. The possibility space
represents the modal content added to the discourse about
properties of the system.

Within this frame, the actualization of a possible property
acquires a rigorous meaning. Let ${\cal L}$ be an orthomodular
lattice, $(W_i)_{i \in I}$ the family of Boolean sublattices of
${\cal L}$ and ${\cal L}^\Diamond$ a modal extension of $\cal L$.
If $f: \Diamond {\cal L} \rightarrow {\bf 2}$ is a Boolean
homomorphism, an actualization compatible with  $f$ is a global
valuation $(v_i: W_i \rightarrow {\bf 2})_{i\in I}$ such that
$v_i\mid W_i \cap \Diamond {\cal L} = f\mid W_i \cap \Diamond
{\cal L} $ for each $i\in I$.

A kind of converse of this possibility of actualizing properties
may be read as an algebraic representation of the Born rule,
something that has no place in the orthomodular lattice alone
(Domenech {\it et al.}, 2006a):

\begin{theo} \label{BORN}
Let ${\cal L}$ be an orthomodular lattice, $W$ a Boolean
sublattice of ${\cal L}$ and $f: W \rightarrow {\bf 2}$  a Boolean
homomorphism. If we consider a modal extension ${\cal L}^\Diamond$
of ${\cal L}$ then there exists a Boolean homomorphism $f^*:
\langle W \cup \Diamond {\cal L} \rangle_{{\cal L}^\Diamond}
\rightarrow {\bf 2} $ such that $f^* \mid W = f$.
\end{theo}

Compatible actualizations represent the passage from possibility
to actuality, they may be regarded as formal constrains when
applying the interpretational rule proposed by Dieks. When taking
into account compatible actualizations from different contexts,
the following KS theorem for modalities can be proved (Domenech
{\it et al.}, 2006a):

\begin{theo}\label{ksm}
Let $\cal L$ be an orthomodular lattice. Then $\cal L$ admits a
global valuation iff for each possibility space there exists a
Boolean homomorphism  $f: \Diamond {\cal L} \rightarrow {\bf 2}$
that admits  a compatible actualization.\qed
\end{theo}

\noindent This theorem shows that no enrichment of the
orthomodular lattice with modal propositions allows to circumvent
the contextual character of the quantum language. This is why we
have called \ref{ksm} the modal Kochen-Specker (MKS) theorem.

This may also be seen from a topological point of view. Let us
consider ${\cal L}^\Diamond$, the modal extension of ${\cal L}$.
Then, the spectral sheaf $p$ defined in section 2  is a subsheaf
of $p_{{\cal L}^\Diamond}$. In this case we refer to $p_{{\cal
L}^\Diamond}$ as a {\it modal extension} of $p$. It is clear that
local sections of $p$ can be seen as local sections of $p_{{\cal
L}^\Diamond}$. Then we define the set $Sec (\Diamond {\cal L}) =
\{\nu:(\Diamond {\cal L}] \rightarrow E_{{\cal L}^\Diamond}: \nu
\hspace{0.2cm} {\rm is \ principal \ section \ of} \hspace{0.2cm}
p_{L^\Diamond} \}$. Since $\Diamond {\cal L}$ is a Boolean
algebra, it is a subdirect product of $\bf 2$. Thus, it always
exists a Boolean homomorphism $f:\Diamond {\cal L} \rightarrow 2$,
resulting $Sec (\Diamond {\cal L}) \not= \emptyset$. From a
physical point of view, $Sec (\Diamond {\cal L})$ represents all
physical properties as possible properties. The fact that $Sec
(\Diamond {\cal L}) \not= \emptyset$ shows that, {\it in the frame
of possibility}, one may talk simultaneously about all physical
properties. The (always possible) choice of a context in which any
possible property pertaining to this context can be considered as
an actual one, may be formalized in the following way: let $\cal
L$ be an orthomodular lattice, $W$ a Boolean sublattice of $\cal
L$, $q\in W$ and ${\cal L}^{\Diamond}$ be a modal extension of
$\cal L$. If $\nu \in Sec (\Diamond {\cal L}) $ such that
$\nu(\Diamond q) = 1$ then an actualization of $q$ compatible with
$\nu$ is an extension $\nu':U \rightarrow E_{{\cal L}^\Diamond}$
such that $(\langle W \cup \Diamond {\cal L} \rangle_{L^\Diamond}]
\in U$. Then we may prove (Domenech {\it et al.}, 2006b) that, if
$\nu \in Sec (\Diamond {\cal L}) $ such that $\nu(\Diamond q) =
1$, then there exists an actualization of $q$ compatible with
$\nu$. It is also possible to represent the Born rule in terms of
continuous local sections of sheaves: let $\cal L$ be an
orthomodular lattice, $W$ a Boolean sublattice of $\cal L$, and
$\nu:(W]\rightarrow E_{\cal L} $ a principal local section. If we
consider a modal extension ${{\cal L}^\Diamond}$ of $\cal L$ then
there exists an extension $\nu':U \rightarrow E_{{\cal
L}^\Diamond}$ such that $\langle W \cup \Diamond {\cal L}
\rangle_{{\cal L}^\Diamond} \in U$. This rule quantifies
possibilities from a chosen spectral algebra and is a kind of the
converse of the possibility of actualizing properties to which we
have referred before. Now, an actualization compatible with $\nu$
is a global section $\tau: {\cal W}_{\cal L} \rightarrow E_{\cal
L}$ of $p_{\cal L}$ such that $\tau(W \cap \Diamond {\cal L}) =
\nu(W \cap \Diamond {\cal L})$. Thus we may state the following
theorem, the topological version of the MKS theorem:

\begin{theo}\label{ksm2}
Let $\cal L$ be an orthomodular lattice. Then $p_{\cal L}$ is a
global section $\tau$ iff for each modal extension ${{\cal
L}^\Diamond}$ there exists  $\nu \in Sec (\Diamond {\cal L})$ such
that $\tau$ is a compatible actualization of $\nu$.\qed
\end{theo}

In view of this theorem, since any global section of the spectral
sheaf is a compatible actualization of a local one belonging to
$Sec (\Diamond {\cal L})$, a global actualization that would
correspond to a family of compatible valuations is prohibited.
Thus, the theorem provides the same conclusion of the MKS theorem,
but now from a topological point of view.\\

The scope of the MKS theorem and its topological version
\ref{ksm2} supersedes the usual KS-type theorems, which only refer
to the actual values of physical properties. It allows to take
into account also possible properties which enlarge the expressive
power of the discourse. In spite of the fact that at first sight
it may be thought that referring to possibility could help to
circumvent contextuality, allowing to refer to physical properties
belonging to the system in an objective way that resembles the
classical picture, our theorems show that this cannot be
consistently maintained.

\section*{Discussion}

Our interest in KS-type theorems in relation to modalities steams
from the fact that quantum mechanics has referred, from its very
beginning, not only to actuality but also to possibility. In this
sense, our problem has been to clarify the relation between
contextuality and modality in quantum mechanics. We have chosen a
logical approach because, though the main contribution of a
logical calculus is actually rather technical, it makes visible
the structure in which propositions lie  and provides bounds to a
consistent discourse. It is a powerful tool when dealing with
non-classical propositions.

We have shown that the addition of modalities to the discourse
about the properties of a quantum system genuinely enlarges its
expressive power. More precisely, the usual orthomodular
propositional structure ${\cal L}$  that does not contain modal
elements is embedded in a Boolean saturated orthomodular lattice
${\cal L^{\Diamond}}$, its modal extension, to obtain a common
frame. But in view of MKS theorem, a global actualization that
would correspond to a family of compatible valuations is
prohibited. Thus, contextuality remains a central feature of
quantum systems even when possibilities are taken into account by
enriching the structure with modal propositions.

The MKS theorem imposes clear constrains in order to put forward
an interpretation in terms of objective probabilities, such as
those proposed by Popper in terms of propensities. If one is
willing to have an interpretation of probability in terms of
objective chance, i.e. a certain measure over possible worlds,
there must be some realistic support, a sense in which these
possible worlds are real. As noted by Schr\"{o}dinger, ``{\it A
probabilistic assertion presupposes the full reality of its
subject.}'' Our modal extension provides exactly the adequate
framework to think in terms of possible worlds, but at the same
time, through our MKS theorem, it precludes the interpretation of
possibility in terms of objective probability.

Finally, we would like to remark that our formalism also provides
a formal meaning in an algebraic frame to the Born rule, something
that has been discussed recently by D. Dieks (2007) in relation to
the possible derivation of a preferred probability measure.

Modal interpretations still have to present a consistent image of
the theory. In words of van Fraassen, the theory should tell us
how the world is like if the theory is true. We believe that in
order to do so, a central feature that needs to be further
investigated is the interpretation of modalities in quantum
mechanics. Regarding the idea of introducing potentiality in
quantum mechanics (see, for example, Karakostas, 2004 and Smets,
2005), it remains for us of great interest to study the constrains
under which such interpretation may be applied. It is important to
stress that the idea of interpreting modalities in terms of
ontological potentiality (de Ronde, 2005) may allow to escape the
constrains imposed by the MKS theorem, as will be discussed in a
forthcoming paper (Domenech {\it et al.}, 2007a).

\section*{Acknowledgements}

This article was presented in the 33rd Annual Philosophy of
Science Conference, Inter-University Centre Dubrovnik (April
10-14, 2006, Dubrovnik, Croatia). We are specially grateful to
Itamar Pitowsky for pointing out the consequences of the MKS
theorem regarding the interpretation of objective probabilities.
We thank Karin Verelst for her careful reading of an earlier
draft. This work was partially supported by the following grants:
PICT 04-17687 (ANPCyT), PIP N$^o$ 1478/01 (CONICET), UBACyT N$^o$
X081 and X204 and Projects of the Fund for Scientific Research
Flanders G.0362.03 and G.0452.04.

\end{document}